\RequirePackage{fixltx2e}
\documentclass[article,A4,twocolumn,amsmath,superscriptaddress,showpacs,floatfix]{revtex4-1}

\usepackage{siunitx}
\usepackage{graphicx}
\usepackage{color}
\usepackage{soul}

\usepackage{siunitx}

\begin{document}

\title{Spin-photon interface and spin-controlled photon switching in a nanobeam waveguide}
\author{Alisa Javadi}
\affiliation{Niels Bohr Institute, University of Copenhagen, Blegdamsvej 17, DK-2100 Copenhagen, Denmark}
\author{Dapeng Ding}
\affiliation{Niels Bohr Institute, University of Copenhagen, Blegdamsvej 17, DK-2100 Copenhagen, Denmark}
\author{Martin Hayhurst Appel}
\affiliation{Niels Bohr Institute, University of Copenhagen, Blegdamsvej 17, DK-2100 Copenhagen, Denmark}
\author{Sahand Mahmoodian}
\affiliation{Niels Bohr Institute, University of Copenhagen, Blegdamsvej 17, DK-2100 Copenhagen, Denmark}
\author{Matthias C. L\"{o}bl}
\affiliation{Department of Physics, University of Basel, Klingelbergstrasse 82, CH-4056 Basel, Switzerland}
\author{Immo S\"{o}llner}
\affiliation{Department of Physics, University of Basel, Klingelbergstrasse 82, CH-4056 Basel, Switzerland}
\author{R{\"u}diger Schott}
\affiliation{Lehrstuhl f{\"u}r Angewandte Festk{\"o}rperphysik, Ruhr-Universit{\"a}t Bochum, Universit{\"a}tsstrasse 150, D-44780 Bochum, Germany}
\author{Camille Papon}
\affiliation{Niels Bohr Institute, University of Copenhagen, Blegdamsvej 17, DK-2100 Copenhagen, Denmark}
\author{Tommaso Pregnolato}
\affiliation{Niels Bohr Institute, University of Copenhagen, Blegdamsvej 17, DK-2100 Copenhagen, Denmark}
\author{S{\o}ren Stobbe}
\affiliation{Niels Bohr Institute, University of Copenhagen, Blegdamsvej 17, DK-2100 Copenhagen, Denmark}
\author{Leonardo Midolo}
\affiliation{Niels Bohr Institute, University of Copenhagen, Blegdamsvej 17, DK-2100 Copenhagen, Denmark}
\author{Tim Schr\"{o}der}
\affiliation{Niels Bohr Institute, University of Copenhagen, Blegdamsvej 17, DK-2100 Copenhagen, Denmark}
\author{Andreas D.~Wieck}
\affiliation{Lehrstuhl f{\"u}r Angewandte Festk{\"o}rperphysik, Ruhr-Universit{\"a}t Bochum, Universit{\"a}tsstrasse 150, D-44780 Bochum, Germany}
\author{Arne Ludwig}
\affiliation{Lehrstuhl f{\"u}r Angewandte Festk{\"o}rperphysik, Ruhr-Universit{\"a}t Bochum, Universit{\"a}tsstrasse 150, D-44780 Bochum, Germany}
\author{Richard J.~Warburton}
\affiliation{Department of Physics, University of Basel, Klingelbergstrasse 82, CH-4056 Basel, Switzerland}
\author{Peter Lodahl}\email{lodahl@nbi.ku.dk}
\affiliation{Niels Bohr Institute, University of Copenhagen, Blegdamsvej 17, DK-2100 Copenhagen, Denmark}

\date{\today}



\maketitle

\textbf{Access to the electron spin is at the heart of many protocols for integrated and distributed quantum-information processing \cite{Kimble2008QuantumInternet,Duan2004PRL,Meter2010IJQ,Gao2015NPHOT}. For instance, interfacing the spin-state of an electron and a photon can be utilized to perform quantum gates between photons \cite{Duan2004PRL,Hacker2016Nature} or to entangle remote spin states \cite{Delteil2016NPHYS,Hucul2014NPHYS,Sipahigil2016Science,Cirac1997PRL}. Ultimately, a quantum network of entangled spins constitutes a new paradigm in quantum optics \cite{Kimble2008QuantumInternet}. Towards this goal, an integrated spin-photon interface would be a major leap forward. Here we demonstrate an efficient and optically programmable interface between the spin of an electron in a quantum dot and photons in a nanophotonic waveguide. The spin can be deterministically prepared with a fidelity of 96\%. Subsequently the system is used to implement a ``single-spin photonic switch'', where the spin state of the electron directs the flow of photons through the waveguide. The spin-photon interface may enable on-chip photon-photon gates \cite{Duan2004PRL}, single-photon transistors \cite{Chang2007NPHYS}, and efficient photonic cluster state generation \cite{Lindner2009PRL}.
}

Solid-state quantum emitters embedded in planar nanostructures offer a scalable route to integrated light-matter interfaces \cite{Sipahigil2016Science,Faez2014PRL,Lodahl2015RMP}. Among these emitters, InGaAs quantum dots are arguably the most developed platform. Quantum dots have been integrated in various nanostructures with near-unity coupling efficiencies \cite{Lund-Hansen2008PRL,Arcari2014PRL,Claudon2010NPHOT,Laucht2012PRX,Makhonin2014NLET}. Such a high coupling efficiency has enabled near deterministic and indistinguishable single-photon sources \cite{Ding2016PRL,Kirsanske2017Arxiv}, as well as single-photon level nonlinearities \cite{Javadi2015NCOM,Santis2017NNANO,Bennett2016NNANO}. Furthermore, chiral light-matter interaction leading to directional photon emission and scattering \cite{Luxmoore2013PRL,Sollner2015NNANO,Coles2016NCOM} has opened new prospects for integrated quantum-information processing \cite{Lodahl2017Nature}.

\begin{figure*}[ht!]
  \includegraphics[width=180 mm]{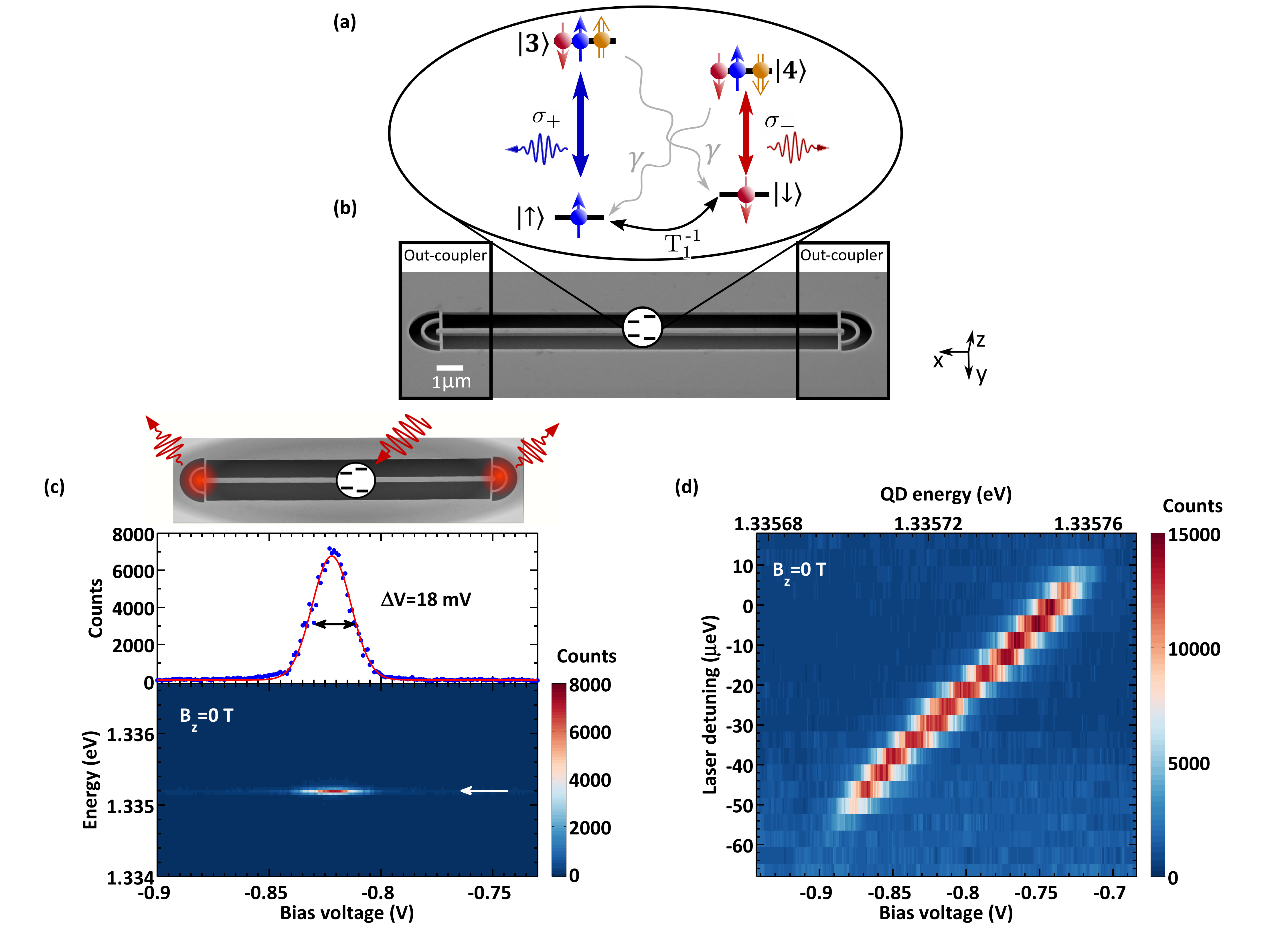}\\
  \caption{\textbf{Resonant spectroscopy of a negatively-charged quantum dot in a nanophotonic waveguide.} (a) Level structure of a negatively-charged quantum dot. The $\left|\uparrow\right>$ and $\left|\downarrow\right>$ ground states correspond to spin quantum numbers $m_j=+\frac{1}{2}$ and $m_j=-\frac{1}{2}$. A magnetic field ($\textrm{B}_z$) along the growth axis (Faraday geometry) splits the two ground states. The diagonal transitions are inhibited due to optical selection rules, but weakly allowed due to heavy-hole light-hole mixing. $\gamma$ is spontaneous decay rate through the diagonal transition, $\textrm{T}_1$ is the ground state lifetime, and $\sigma_+$ ($\sigma_-$) indicate a right- (left-) hand circularly polarized dipole transition. (b) Scanning electron micrograph of a nanobeam waveguide. The quantum dot is located close to the centre of the waveguide and an electric field is applied across the quantum dot using a diode structure. (c) Emission spectrum of $\textrm{X}^-$ under resonant excitation at $\textrm{B}_z=\SI{0}{T}$. The quantum dot is excited from the top of the waveguide with x-polarized light at an energy indicated by the white arrow. The emission is collected from the gratings and the quantum dot is tuned through the resonance by varying the bias voltage. The top panel shows a line-cut along the laser energy. The solid curve is a Gaussian fit with a line-width (FWHM) of $\Delta \textrm{V}=\SI{18}{mV}$, corresponding to $\Delta \textrm{E} = \SI{7.4}{\mu eV}$. (d) Plateau map at $\textrm{B}_z=\SI{0}{T}$. Resonance fluorescence from the $\textrm{X}^-$ exciton, which is plotted as a function of the laser energy and the applied bias voltage. The top axis shows the energy of the $\textrm{X}^-$ transition.}\label{fig1}
\end{figure*}

Significant progress has been made on coherent control of the spin state of electrons and holes in quantum dots \cite{Kroutvar2004Nature,Gerardot2008Nature,Atature2006Science,Warburton2013NMAT,Press2010NPHOT,Stockill2016NCOM}, spin-photon entanglement \cite{De2012Nature,Gao2012Nature}, and spin-spin entanglement \cite{Delteil2016NPHYS}. However, coupling of the spin of quantum dots to planar nanostructures is so far largely unexplored, and has only been studied with in-plane magnetic fields (Voigt geometry) \cite{Carter2013NPHOT,Sun2016NNANO}, and with probabilistically-charged quantum dots \cite{Sun2016NNANO}. The first challenge is that the quantum dot needs to be placed in thin optical membranes (typically with a thickness below \SI{200}{nm}) to obtain single-mode operation and therefore efficient light-matter interaction. Achieving deterministic control over the charge state of the quantum dot in thin membranes has been challenging \cite{Pinotsi2001IEEEJQP}, and has only been demonstrated with samples having a large leakage current ($>$ \SI{4}{mA}) running through the sample \cite{Carter2013NPHOT,Sweeney2014NPHOT,Pinotsi2001IEEEJQP}, which may degrade the spin properties of the quantum dot. As a result, quantum dot spin-state preparation has only been achieved with in-plane external magnetic fields \cite{Carter2013NPHOT,Sun2016NNANO}, which do not require long spin lifetimes \cite{Emary2007PRL}. Secondly, many groups have focused on achieving strong coupling between the spin state of the quantum dot and a narrow-linewidth cavity \cite{Carter2013NPHOT,Lagoudakis2013NJP,Sweeney2014NPHOT,Sun2016NNANO}. However, this approach is challenging to implement in spin-photon interfaces since the narrow cavity bandwidth implies that the two different spin states cannot simultaneously be well-coupled. Complete control of the spin state of a quantum dot involves driving and detecting four different optical transitions simultaneously, c.f, Fig. \ref{fig1}a.

Nanophotonic waveguides offer high coupling efficiencies of embedded quantum emitters over a wide spectral bandwidth \cite{Lodahl2015RMP}, and appear to be an ideal platform for realizing efficient spin-photon interfaces. This is the setting referred to as the ``1D quantum emitter'', where the coupling of the emitter to a single optical mode is so pronounced that deterministic photon-emitter coupling may be achieved. In the present work we demonstrate an integrated interface between the spin state of an electron in a quantum dot and the optical mode of a nanobeam waveguide. We use this interface to realize a proof-of-concept optically programmable photon switch, which is controlled by the spin state of the quantum dot. We achieve deterministic charging of the quantum dot at low bias voltages and with very low parasitic leakage current ($\SI{1}{\mu A}$, corresponding to a current density of $\sim\SI{0.13}{\mu A/mm^2}$), which allows us to demonstrate high-fidelity spin state preparation while applying a magnetic field parallel to the growth axis of the quantum dot (Faraday geometry). The spin preparation fidelities approach 96\% for a spin lifetime of $T_1\sim\SI{4}{\mu s}$. We observe that the diagonal decay rate ($\gamma$) is two orders of magnitude smaller than the vertical decay rate ($\Gamma$), $\Gamma\simeq100\gamma$. Together with the long spin lifetime, ${\textrm{T}_1}\simeq50\gamma^{-1}$, this is the favourable condition for realizing high-fidelity spin state preparation \cite{Warburton2013NMAT,Gao2015NPHOT} as required for photon-photon gates and single-photon transistors \cite{Duan2004PRL,Chang2007NPHYS}. Finally, we show that the quantum dot modifies the transmission through the waveguide, and, by deterministically preparing the spin state of the electron by fast optical control, we can switch the transmission of the waveguide. The observed spin-dependent transmission can be utilized as a programmable switch for routing single photons on a chip. Contrary to operation in the Voigt geometry, the Faraday configuration features chiral light-matter interaction and may enable nonreciprocal light transport \cite{Sollner2015NNANO}, which can be employed for novel quantum photonics devices such as on-chip single-photon circulators \cite{Scheucher2016Science}, integrated isolators \cite{Sayrin2015PRX}, and building blocks for distributed quantum networks \cite{Mahmoodian2017PRL}.

\subsection*{Level structure}
Figure \ref{fig1}a shows the level structure of a negatively charged quantum dot under influence of a static magnetic field along the growth axis ($\textrm{B}_z$) \cite{Warburton2013NMAT}. In the ground state, the spin of the trapped electron is oriented parallel or anti-parallel to the growth axis, which are labeled as spin-up ($\left|\uparrow\right>$) and spin-down states ($\left|\downarrow\right>$), respectively.
The two excited states are negatively-charged excitons ($\textrm{X}^-$), and consist of two spin-paired electrons and a single hole. The vertical transitions are circularly polarized ($\sigma_+$ and $\sigma_-$).
The diagonal transitions are weakly allowed due to the in-plane Overhauser field and heavy-hole light-hole mixing in the valence band of the quantum dot, with the latter being the dominant mechanism at higher magnetic fields \cite{Dreiser2008PRB}.

\subsection*{Sample design and characterization}
Figure \ref{fig1}b shows a scanning electron micrograph (SEM) of a nanobeam waveguide, terminated with grating out-couplers at the two ends. A layer of InGaAs quantum dots is positioned in the centre of the membrane inside a P-I-N-I-N diode grown along the $z$-direction \cite{Lobl2017inprep}. The design of the diode facilitates deterministic charging of the quantum dot as well as tuning of its energy levels by applying a bias voltage, see Supplementary Information for details.

The colour plot in Fig.~\ref{fig1}c shows a spectrum of the emission from $\textrm{X}^-$ under weak resonant excitation at $\textrm{B}_z=\SI{0}{T}$. The laser is fixed at \SI{1.3352}{eV} and the energy of the quantum dot transition is tuned by changing the bias voltage. The top panel shows a line-cut through the laser energy. The linewidth of the resonance is \SI{7.4}{\mu eV} ($\sim 9 \times$ the natural linewidth), where the broadening is attributed to charge noise from the environment of the quantum dot \cite{Kuhlmann2013NPHYS,Nguyen2013PRB}, although not a fundamental limitation in nanobeam waveguides where narrow linewidth quantum dots were recently observed \cite{Kirsanske2017Arxiv}. Figure \ref{fig1}d shows a plateau map of the resonance fluorescence from the $\textrm{X}^-$ transition as a function of excitation laser energy and the bias voltage. The relevant transition energy range is between \SI{1.335696}{eV} and \SI{1.335760}{eV}, where the quantum dot is charged with a single electron in the ground state. Below this plateau region the quantum dot is empty and above the quantum dot is charged with two electrons and hence the fluorescence from $\textrm{X}^-$ vanishes \cite{Warburton2000Nature,hogele2004voltage}. In the centre of the plateau, the single-electron charged state of the quantum dot is a stable state, and the spin state of the electron is only influenced by second-order processes such as co-tunneling with the back contact \cite{Smith2005PRL,Dreiser2008PRB}, or by Auger recombination \cite{Kurzmann2016NLET}.

\subsection*{Optical spin state preparation}
\begin{figure*}[t!]
  \includegraphics[width=180 mm]{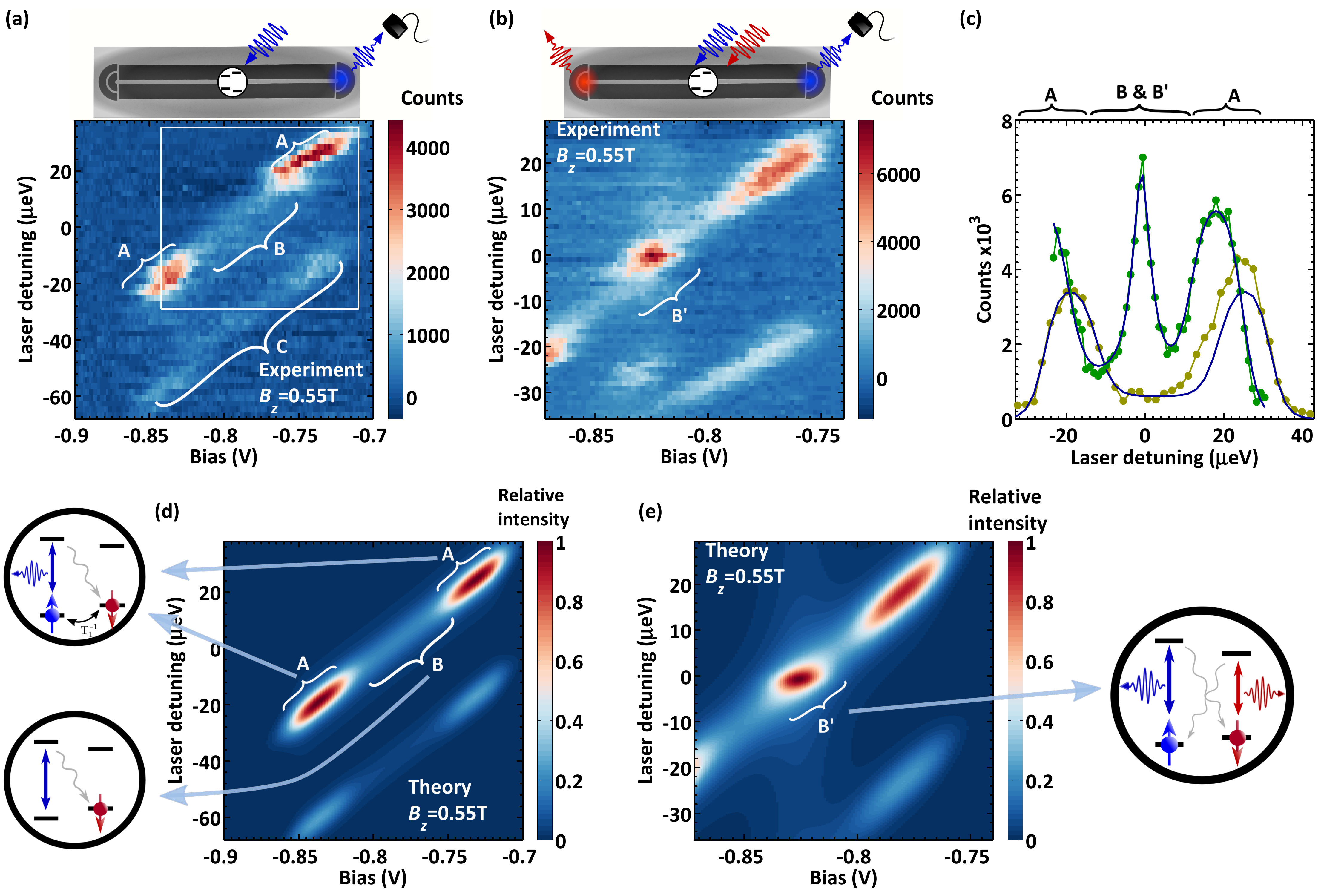}\\
  \caption{\textbf{Quantum dot spin preparation in a nanophotonic waveguide.} (a) Plateau map of the $\textrm{X}^-$ transitions at $\textrm{B}_z=\SI{0.55}{T}$. The plateau regions A and B correspond to fluorescence from the blue transition in Fig. \ref{fig1}a, while the region marked as C corresponds to emission from the red transition. The two plateaus have different intensities due to chiral light-matter interaction. The central part of the plateau is dim, due to optical spin-pumping from the $\left|\uparrow\right>$ state to the $\left|\downarrow\right>$ state and vice versa. At the edges of the plateau (region A), spin pumping is suppressed due to co-tunneling between the electron in the quantum dot and the Fermi-sea in the back contact. (b) Same as (a) with a second laser fixed at \SI{-40}{\mu eV}, which corresponds to the centre of the $\left|\downarrow\right>$ plateau. The bright spot in the centre ($\textrm{B}'$) is caused by simultaneous excitation of the blue and red transitions, when the first laser is tuned to the centre of the plateau of the blue transition. The plotted range corresponds to the white box in (a). The resonance voltages are shifted by \SI{50}{mV} due to charge screening effects induced by the second laser. (c) Line-cuts through the top plateau in part (a) and part (b), coloured as yellow and green, respectively. The solid curves are theoretical fits to the data. (d) and (e) Theoretical models of the plateau maps in (a) and (b). See the Supplementary Information for details of the modeling. By modeling the data in part (a), we extract the spin life time $\textrm{T}_1=\SI{3.8}{\mu s}\pm\SI{1.2}{\mu s}$.}\label{fig2}
\end{figure*}

To prepare two optically accessible spin ground states with an energy difference larger than the transition linewidth, we lift the degeneracy by inducing a Zeeman shift at $\textrm{B}_z=\SI{.55}{T}$. We probe the $\left|\uparrow\right>$ and $\left|\downarrow\right>$ states through resonance fluorescence from the $\textrm{X}^-$ exciton. Figure \ref{fig2}a shows the plateau map of the $\textrm{X}^-$ exciton. The emission plateau (regions A and B) originates from the high energy (blue) transition of the negatively-charged exciton, while region C corresponds to emission from the low energy (red) transition. We note that the blue transition is $\sim4$ times brighter than the red transition, which is due to chiral light-matter interaction \cite{Sollner2015NNANO,Coles2016NCOM,Lodahl2017Nature}.
At the central part of the plateau, region B, optical spin-pumping takes place and the emission from $\textrm{X}^-$ is suppressed due to spin-non-conserving diagonal transitions. At the edges of the plateaus, region A, the electron is strongly coupled to the Fermi-sea in the back contact and its spin is randomized over short time scales ($\sim$ 10s of nano-seconds) \cite{Smith2005PRL,Dreiser2008PRB} which hinders spin-pumping. At these points, the fluorescence is $\sim6-7$ times brighter than in the centre, cf. line-cut data in Fig. \ref{fig2}c. By comparing the resonance fluorescence intensity at the edge of the plateau with the emission at the centre of the plateau, we extract a spin preparation fidelity $\left<\downarrow|\rho|\downarrow\right> \ge 86\%$, where $\rho$ is the density matrix of the prepared state. At higher magnetic fields we achieve fidelities up to $96\%$, see Supplementary Information for details.

To confirm optical spin pumping, we perform a two-colour resonance fluorescence experiment \cite{Kroner2008PRB}, where one laser is fixed at the centre of the plateau of the red transition, while the frequency of the second laser is scanned. Figure \ref{fig2}b shows the two-colour plateau map of the $\textrm{X}^-$ exciton. When the two lasers are on resonance with the blue and red transitions simultaneously, $\textrm{B}'$, they cancel each others' spin-pumping effect and the resonance fluorescence from the quantum dot is recovered.
Figure \ref{fig2}d and \ref{fig2}e are the theoretical models of the data, see Supplementary Information for details. The experimental behavior is quantitatively described by the theory, for $\textrm{T}_1=\SI{3.8}{\mu s}\pm\SI{1.2}{\mu s}$. We also extract $\gamma=\SI{13}{MHz}\pm\SI{1}{MHz}$ using time resolved measurements, the details are described in the Supplementary Information. $\textrm{T}_1$ may also be extracted directly by pump-delay-probe experiments where we obtain $\textrm{T}_1=\SI{4.3}{\mu s}\pm\SI{0.2}{\mu s}$, in very good agreement with the parameters extracted from modeling the data in Fig. \ref{fig2}a. We emphasize that the observed spin lifetime in the photonic nanostructure is longer than the longest spin coherence time reported with quantum dots \cite{Wust2016NNANO} even after implementing spin-echo techniques \cite{Press2010NPHOT}, hence the spin relaxation will not limit any processes which depend on spin coherence.

\subsection*{Spin-controlled photon switching }
\begin{figure*}[t]
  \includegraphics[width=180 mm]{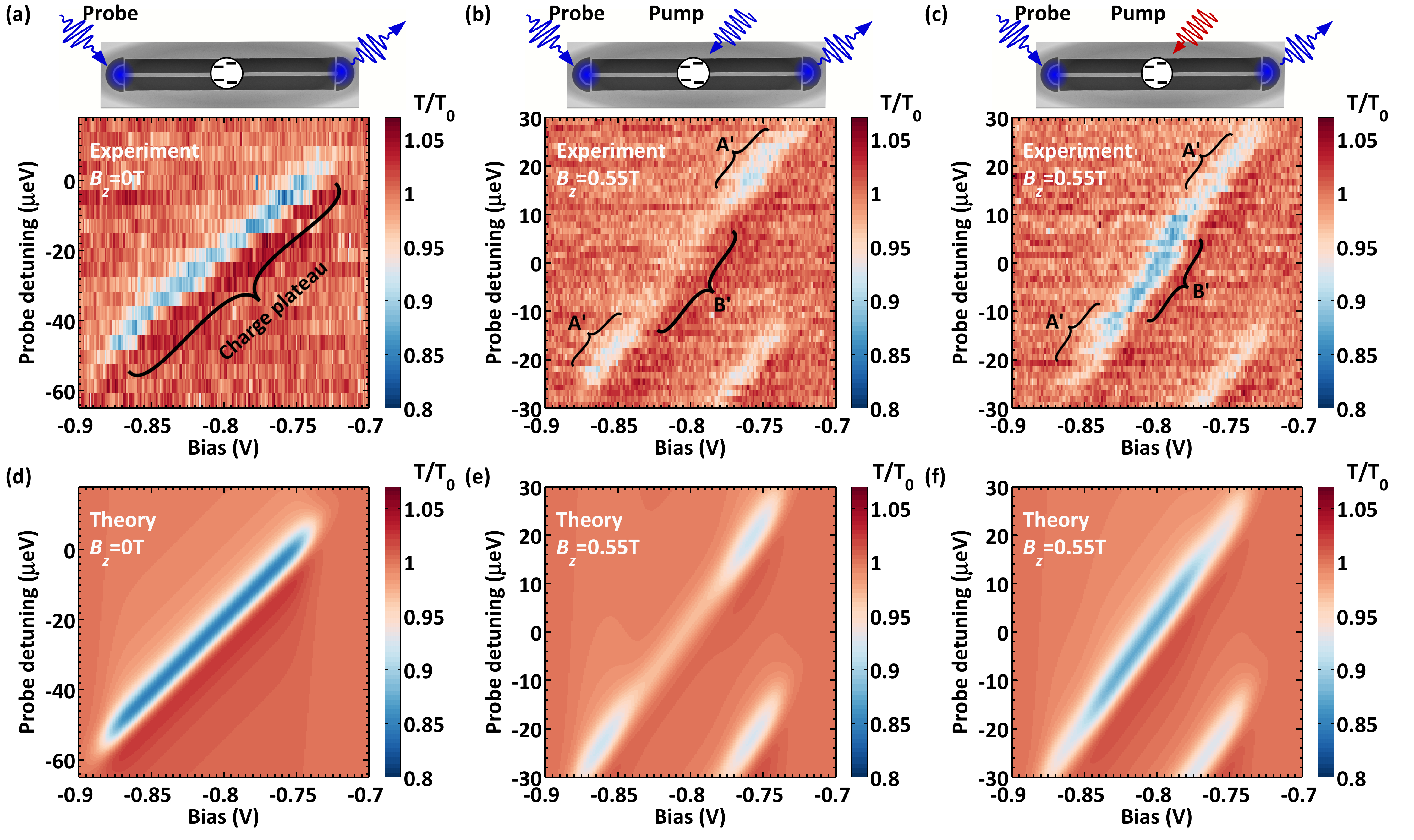}\\
  \caption{\textbf{Spin-controlled resonant transmission through the nanophotonic waveguide.}(a) Measurement of the transmission through the waveguide as a function of the probe pulse energy and the bias voltage at $\textrm{B}_z=\SI{0}{T}$. A weak probe pulse is launched into the waveguide through the grating on the left hand side and the transmission is monitored. When the probe is on resonance with the quantum dot transition, the scattering from the quantum dot reduces the transmission of the probe. The probe power is set an order of magnitude below the saturation power of the quantum dot. (b) and (c) Pump-probe measurements at $\textrm{B}_z=\SI{0.55}{T}$, while the probe is swept along the charge plateau of the blue transition. A resonant pump laser launched from the top of the waveguide prepares the spin state of the electron during the pump cycle. The detuning between the probe and pump is fixed to \SI{0}{\mu eV} for (b), and \SI{40}{\mu eV} for (c). In (b) the spin state of the electron is prepared in the $\left|\downarrow\right>$ state during the pump cycle. As a result, the blue transition is turned off, and the transmission of the probe is maximized. $\textrm{A}'$ indicates the edges of the charge plateau where co-tunneling is strong. In (c), the pump laser prepares the electron spin in $\left|\uparrow\right>$ state, and the blue transition reduces the transmission of the probe. (d)-(f) Theoretical models of the experiments in (a)-(c). See the Supplementary information for the details of the models.}\label{fig3}
\end{figure*}

A quantum emitter coupled to a single optical mode can modify the light transmission properties significantly \cite{Javadi2015NCOM,Shen2005OL,Chang2007NPHYS}. For an efficient coupling this interaction can be sensitive at the single-photon level.
The colour map in Fig. \ref{fig3}a shows the normalized transmission of a weak probe as a function of its energy and the bias voltage of the diode at $\textrm{B}_z=\SI{0}{T}$. When the probe is on resonance with the $\textrm{X}^-$ transition, the transmission of the probe is reduced due to the interaction with the transition \cite{Javadi2015NCOM,Shen2005OL}, where the quantum dot scatters one photon at a time. This is observed as a dip in the plateau map in Fig. \ref{fig3}a. We observe a maximum contrast of 15\% in the transmission, which is mainly limited by the inhomogeneous broadening of the quantum dot transition.

To implement a spin-state dependent interaction between the quantum dot and the waveguide mode, we use optical spin pumping to deterministically prepare the spin state of the quantum dot. A strong laser pulse (pump) with a duration of \SI{1}{\mu s} incident from the top of the waveguide prepares the spin state of the electron. Subsequently, a weak pulse (probe) with a duration of \SI{200}{ns}, coupled in and out via the gratings, probes the single-photon transmission through the waveguide.
Figure \ref{fig3}b shows the normalized transmission through the waveguide while the probe and pump pulses are on resonance. In the centre of the plateau, $\textrm{B}'$, the pump pulse prepares the spin of the electron in $\left|\downarrow\right>$ with a high fidelity. This state is off resonance with the probe and hence the transmission recovers to the level $T_0$, which is the level encountered when the probe is far from resonance of the quantum dot transition. At the edges of the plateau, $\textrm{A}'$, the transmission of the probe is reduced due to inefficient spin state preparation.
Figure \ref{fig3}c shows the transmission of the waveguide while the pump laser is detuned by \SI{-40}{\mu eV} from the probe. In this case, the pump prepares the spin of the electron in the $\left|\uparrow\right>$ state, which is on resonance with the probe. As a result, the probe pulse interacts with the blue transition of the $\textrm{X}^-$ exciton. At the centre of the plateau ($\textrm{B}'$) the transmission of the waveguide is reduced to 0.87, similar to the value found without an external magnetic field.

\begin{figure}[ht!]
  \includegraphics[width=80 mm]{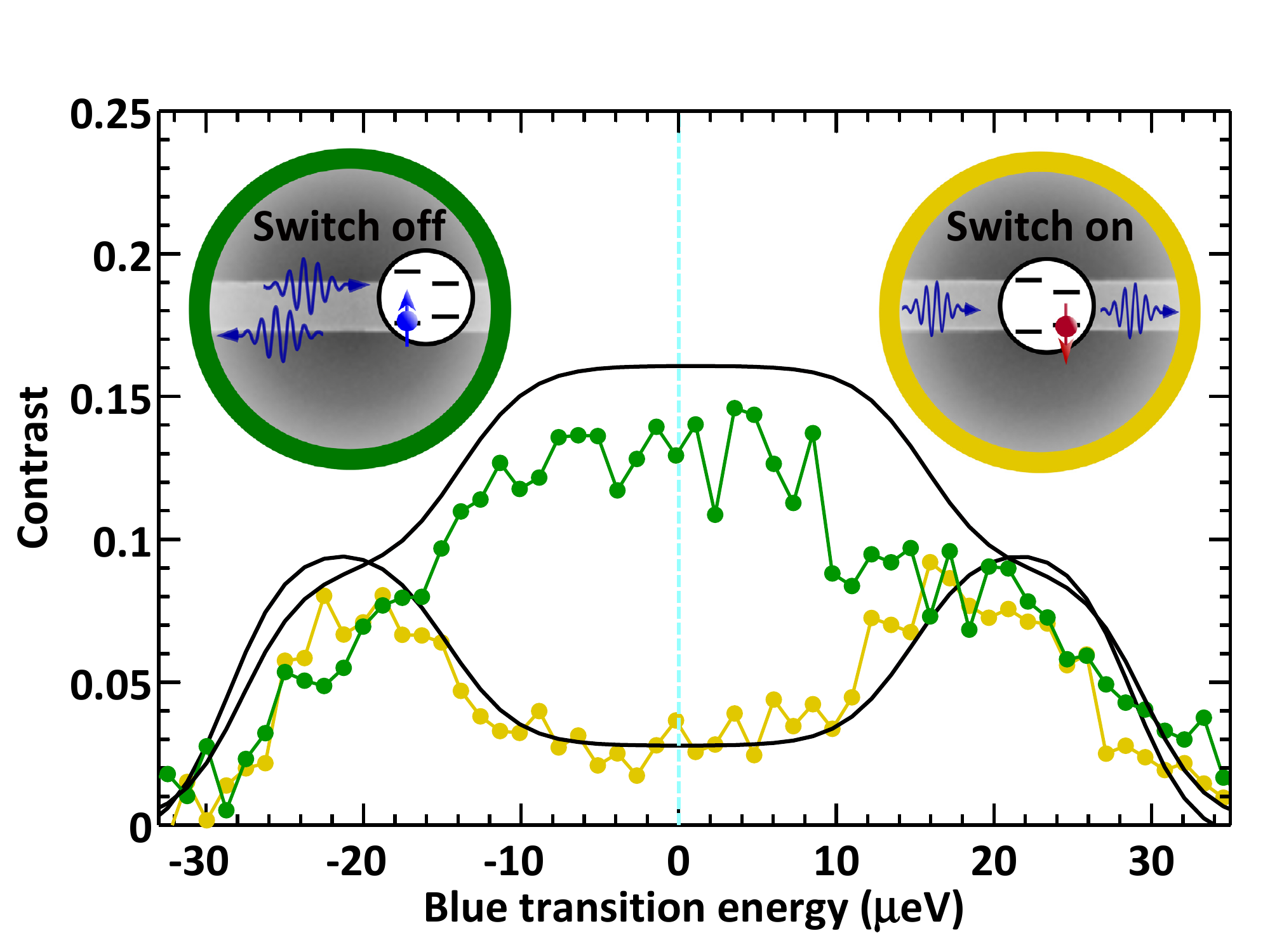}\\
  \caption{\textbf{Spin-controlled switching of the transmission.} Contrast in the transmission of the waveguide as a function of the detuning from the blue transition. The top left insert is a schematic demonstration of the reflection of photons by the quantum dot when the spin of the electron is prepared in the $\left|\uparrow\right\rangle$ state, corresponding to the green data points. The right-hand insert shows the case where the spin of the electron is prepared in $\left|\downarrow\right\rangle$ and the quantum dot does not interact with the probe. From the data in this figure we infer a switching contrast ratio of 4. The data correspond to line-cuts through $\textrm{A}'$ and $\textrm{B}'$ regions in Fig. \ref{fig3}b (green), and Fig. \ref{fig3}c (yellow). The solid curves are theoretical models based on the parameters extracted from the resonance fluorescence experiments (See Supplementary Information)}\label{fig4}
\end{figure}

The ability to prepare a spin deterministically and thereby control the waveguide transmission constitutes a proof-of-concept realization of a switch for single photons. Figure \ref{fig4} shows the contrast of the transmission through the waveguide as a function of the energy detuning. The transmission is switched by tuning of the pump pulse to either the red or the blue transition of the quantum dot. The ON and OFF states of the switch correspond to the case where the spin of the quantum dot is prepared in $\left|\downarrow\right\rangle$ and $\left|\uparrow\right\rangle$, respectively. We observe a switching ratio of more than a factor of 4 between ON and OFF states, which could be improved further by reducing spectral diffusion due to residual charge noise broadening. Ultimately the switch could be operated in a genuine quantum regime if the spin was initially prepared in a coherent superposition state. Such a quantum switch could create a photonic Schr{\"{o}}dinger cat state when applied to a weak coherent state \cite{Chang2007NPHYS}.

It is instructive to benchmark the reported performance of all-optical switching. The pump power on the sample for the data in Fig. \ref{fig4} is around \SI{40}{nW} for a duration of \SI{1}{\mu s}, which corresponds to 40 femto joules per pump cycle. During each switching cycle the quantum dot scatters $\Gamma/\gamma$ photons, which is about 100 photons for the measurements reported here \cite{Chang2007NPHYS}. The average energy to switch one photon is therefore $\sim 0.4$ femto joules per switched photon, which is comparable to the switching energy of cavity-based switches \cite{Fushman2008Science,Volz2012NPHOT}. The energy could be further reduced by up to two orders of magnitude by directly launching the pump pulse inside the waveguide. The switching time in the present work is significantly slower than the above mentioned approaches based on the strong coupling of a quantum dot exciton transition to a cavity. However, the present approach offers the advantage of having access to the quantum memory in the form of the ground state electron spin, which is the prerequisite for many quantum-information applications \cite{Lodahl2017Arxiv}.

We have demonstrated all-optical control of a single electron spin in a quantum dot efficiently coupled to a nanophotonic waveguide. Based on this approach, a single spin controls the flow of photons through the waveguide. The work opens a range of new opportunities in quantum optics for exploiting deterministic photon-spin coupling, e.g., for generating long strings of photonic cluster states \cite{Schwartzaah4758Science,Lindner2009PRL}, a high-fidelity photon-photon gate \cite{Duan2004PRL} or single-photon transistors \cite{Chang2007NPHYS}. Extending to the coupling of two quantum dots would enable to construct a fundamental building block for a distributed photonic quantum network \cite{Mahmoodian2017PRL}. For these potential applications it is favorable to work in the Faraday geometry (external magnetic field along the growth direction) as is the case here, since the operation fidelity scales as $\sim1-\gamma/\Gamma$ \cite{Chang2007NPHYS}, where $\Gamma \gg \gamma$ was reported here. One challenge, however, is the coupling of the electron spin to the noisy nuclear bath leading to electron spin dephasing. Recent work has demonstrated how to significantly reduce the noise on the Overhauser field by feedback control of the nuclear ensemble \cite{Xu2009Nature,Kuhlmann2015NCOMM,Ethier2017Arxiv}.

\begin{acknowledgments}
We acknowledge N. J. Taba, and C. L.~Dree{\ss}en for help during the initial stages of the measurements. We gratefully acknowledge financial support from the European Research Council (ERC Advanced Grant ``SCALE''), Innovation Fund Denmark (Quantum Innovation Center ``Qubiz''), and the Danish Council for Independent Research. IS, MCL \& RJW acknowledge support from SNF (project 200020\_156637) and NCCR QSIT. A.L. and A.D.W. gratefully acknowledge support of BMBF - Q.com-H 16KIS0109 and the DFG - TRR 160. This project has received funding from the European Unions Horizon 2020 research and innovation programme under the Marie Skodowska-Curie grant agreement No. 747866 (EPPIC).

\end{acknowledgments}

\subsection*{Contributions}
A.J., D.D., M.H.A., and T.S. carried out the optical experiment with input from I.S., R.J.W., and P.L.
A.J., M.H.A., and S.M. performed the theory. M.C.L., I.S., A.L., and R.J.W. designed the heterostructure. R.S., A.L., and A.D.W. grew the wafer. C.P., T.P., S.S., and L.M. designed and fabricated the sample. A.J. and P.L. wrote the manuscript with input from all authors.

\bibliography{spin_prep_manu_v6}
\end{document}